 \documentstyle[prd,aps,eqsecnum,floats,epsfig]{revtex}
  \tighten
\begin{document}
 \twocolumn[
 \hsize\textwidth\columnwidth\hsize\csname@twocolumnfalse%
 \endcsname
%
\title{Quark exchange and quark dynamics 
in diffractive electroproduction}
\vspace*{-.9cm}\hspace*{\fill}
\parbox{8.0cm}{
  {Talk given at 
  ``Jefferson Lab Physics \& Instrumentation with 6--12 GeV Beams,'' 
   Newport News, VA.}
   {\hspace*{\fill} \sf FSU-SCRI-98-064} 
} \vspace*{+0.9cm}
%
\author{M.A.~Pichowsky}
\address{Department of Physics \&
	Supercomputer Computations Research Institute, \\
        Florida State University, Tallahassee, FL 32306-4130, USA
	}
\date{June 17, 1998}
\maketitle
\begin{abstract}
A Dyson-Schwinger-based model of pomeron exchange is employed to 
calculate diffractive $\rho$-, $\phi$- and $J/\psi$-meson
electroproduction cross sections.
It is shown that the magnitude of the current-quark mass $m_f$ of the
quark and antiquark inside the produced vector meson determines the onset
of the asymptotic-$q^2$ power-law behavior of the cross section, and
how correlated quark-exchanges are included to provide a complete picture
of the diffractive electroproduction of light vector mesons applicable 
over {\em all} energies and photon momenta $q^2$. 
\medskip
\end{abstract}
 ]
\newcommand{\Fig}[1]{Fig.~\protect\ref{#1}}
\newcommand{\Figs}[1]{Figs.~\protect\ref{#1}}
\newcommand{\Ref}[1]{Ref.~\protect\cite{#1}}
\newcommand{\Eq}[1]{Eq.~(\protect\ref{#1})}
\newcommand{\Eqs}[1]{Eqs.~(\protect\ref{#1})}
\renewcommand{\-}{\!-\!}
\renewcommand{\+}{\!+\!}
\newcommand{\T}{{\bf T}}
\newcommand{\sfrac}[2]{\mbox{$\textstyle \frac{#1}{#2}$}}
\newcommand{\nn}{\nonumber\ }        

\section{Introduction}
\footnotetext[1]{
  The Euclidean metric 
  $\delta_{\mu \nu} = {\rm diag}(1,1,1,1)$ is employed with 
  {\em spacelike} momentum $q_{\mu}$ satisfying 
  $q^{2} = q_{\mu} q_{\mu} > 0 $.
 }
Many aspects of diffractive processes are well described within
Regge theory in terms of ``Pomeron exchange''.
However, since the advent of QCD, the underlying mechanism responsible for
pomeron exchange is typically thought to be multiple-gluon exchange.
This suggests that experimental investigations of diffractive processes
provide a means to study gluon correlations in the nonperturbative,
small-momentum-transfer regime of QCD.
The additional freedom to vary the momentum transfer $q^2$
between electron beam and target provides electron beam facilities with
more leverage to study pomeron exchange than facilities employing
hadron beams.
TJNAF provides a crucial tool with which to probe the
nonperturbative-gluon dynamics underlying pomeron exchange. 

In the following, I briefly discuss the elements of a 
covariant, quantum field theoretic quark-nucleon pomeron-exchange model
that was developed in \Ref{Pichowsky} and describe two applications of
this model to diffractive electroproduction.
The model employs elements from studies of the Dyson-Schwinger equations
of QCD, such as dressed-quark propagators which
incorporate confinement, dynamical chiral symmetry breaking, and have the
correct asymptotic behavior required by perturbative QCD \cite{Burden}.
The model provides an excellent description of diffractive processes
and reveals some aspects of the interplay between perturbative and
nonperturbative QCD in these processes.
After giving a brief account of the role of the current-quark mass
in determining the onset of the asymptotic-$q^2$ behavior of diffractive
electroproduction cross sections, I describe how quark exchanges
(important at low energies) are included into the model, thereby providing
a complete picture of the diffractive electroproduction of light-quark
vector mesons over {\em all} energies and photon-momenta $q^2$. 

\section{Vector meson electroproduction}

\begin{figure}[h]
\begin{center}
\epsfig{figure=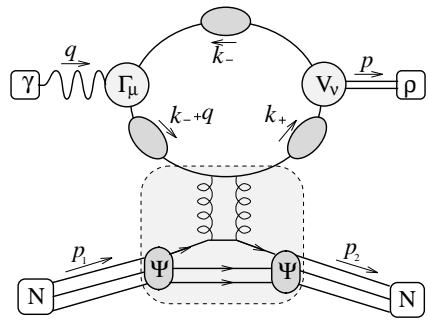,width=5.50cm}
\end{center}
\caption
{
The shaded box represents the quark-nucleon pomeron-exchange interaction
${\cal G}_{\alpha}(W^2,t)$ for $\rho$-meson electroproduction of
\Ref{Pichowsky}.   
Within this box is shown one of many possible
multiple-gluon exchange diagrams that contributes to 
the pomeron-exchange interaction ${\cal G}_{\alpha}(W^2,t)$. 
}
\label{Fig:GRhoGluonX}
\end{figure}

One can study the role of quark dynamics in diffractive
processes using a model in which the high-energy interaction between a
confined-quark and an on-shell nucleon is given in terms of a
pomeron-exchange interaction.   
In principle, this might be calculated in terms of a multiple-gluon
exchange like that shown for $\rho$-meson electroproduction in 
\Fig{Fig:GRhoGluonX}.  
However, this is unnecessary for the purposes of the present discussion.

The $\rho$-meson electroproduction current matrix element
obtained from the pomeron-exchange model of \Ref{Pichowsky}
is\footnotemark[1]
\begin{eqnarray}
\lefteqn{
\langle p \lambda_{\rho} ; p_2 m^{\prime} | J_{\mu}(q) 
| p_1 m \rangle = 
} 
\nn\ \\
& &  2 \; t_{\mu \alpha \nu}(q,p) \;
\varepsilon_{\nu}(p;{\lambda_{\rho}})  \; 
\bar{u}_{m^\prime}(p_2) {\cal G}_{\alpha}(W^2,t) u_m(p_1) 
,
\label{Current}
\end{eqnarray}
where $u_m(p_1)$ and $\bar{u}_{m^\prime}(p_2)$ are, respectively, the
spinors for the incoming and outgoing nucleon, 
$\varepsilon_{\nu}(p;{\lambda_{\rho}})$ is the polarization vector of
the $\rho$ meson, 
${\cal G}_{\alpha}(W^2,t)$ (represented by a shaded box in
\Fig{Fig:GRhoGluonX}) is the model quark-nucleon pomeron-exchange interaction, 
$t_{\mu \alpha \nu}(q,p)$ is the photon-$\rho$-meson transition amplitude
(represented by the quark loop in \Fig{Fig:GRhoGluonX}).
The quark-nucleon interaction ${\cal G}_{\alpha}(s,t)$ is given in terms
of four parameters: the slope and intercept of the
pomeron-exchange Regge trajectory and two coupling constants $\beta_u$ and
$\beta_s$  for $u$- (or $d$-) and $s$-quarks.
Their values are determined in an application to $\pi N$ and $KN$ elastic
scattering \cite{Pichowsky}.   

The photon-$\rho$-meson transition amplitude is given by 
\begin{eqnarray}
\lefteqn{
t_{\mu \alpha \nu}(q,p) = \beta_u N_c e_0 \;
{\rm tr} \int \!\! \frac{{\rm d}^4k}{(2\pi)^4}
\; S_u(k_{--}) 
}
\nonumber \\
& & \times  
\Gamma_{\mu}(k_{--},k_{+-}) 
S_u(k_{+-})  \gamma_{\alpha}  S_u(k_{-+})   
V_{\nu}(k\-\sfrac{1}{2}q;p)
\label{tman} 
,
\end{eqnarray}
where 
$N_c = 3$, $e_0 = \sqrt{4 \pi \alpha_{em}}$ is the elementary charge, 
$S_f(k)$ is the quark propagator of flavor $f$, 
$\Gamma_{\mu}(k,k')$ is the quark-photon vertex, 
$V_{\nu}(k;p)$ is the $\rho$-meson Bethe-Salpeter amplitude, 
$\gamma_\alpha$ is a Dirac matrix, and the trace is over Dirac indices.

Evaluation of $t_{\mu \alpha \nu}(q,p)$ requires explicit forms for the
Schwinger functions $S_f(k)$, $\Gamma_{\mu}(k,k')$, and $V_{\nu}(k,p)$ in
\Eq{tman}.   These are taken from phenomenological studies of hadron
observables based on the Dyson-Schwinger equations of QCD.  

The most general quark-photon vertex 
$\Gamma_{\mu}(k,k^\prime )$ = 
$\Gamma^{\rm T}_{\mu}(k,k^{\prime}) + 
\Gamma^{\rm BC}_{\mu}(k,k^{\prime})$, 
where $\Gamma^{\rm T}_{\mu}(k,k^{\prime})$ is transverse to the photon
momentum $q_{\mu} = (k - k^{\prime})_{\mu}$ and 
$\Gamma^{\rm BC}_{\mu}(k,k^{\prime})$ is {\em completely determined} from
the Ward-Takahashi identity and the dressed quark propagator $S_f(k)$.
Phenomenological studies find that the transverse contributions to
the vertex in $\Gamma^{\rm T}_{\mu}(k,k^{\prime})$ are unimportant for the 
calculation of observables for spacelike photon momenta.
Hence, it is reasonable in the present study to neglect this term
and take the dressed quark-photon vertex as 
$\Gamma^{\rm BC}_{\mu}(k,k^{\prime})$.  This entails a parameter-free
nonperturbatively dressed quark-photon vertex that is consistent with
the dressed-quark propagator $S_f(k)$.

In a general covariant gauge, the dressed-quark propagator is written as
$S_f(k) = - i \gamma \cdot k \sigma_V^f(k^2) + \sigma_S^f(k^2)$.
Numerical studies \cite{Maris} of the Dyson-Schwinger equations find that
the essential features of the dressed-quark propagator are well
represented by a simple parametrization: 
\begin{eqnarray}
  \bar{\sigma}^f_S(x) & = & 
    \left(b^f_0 + b^f_2 {\cal F}[ \epsilon x]  \right)
    {\cal F}[b^f_1 x] {\cal F}[b^f_3 x]
      \nonumber\\    
   & &  
    + 2 \bar{m}_f {\cal F}\big[ 2(x+\bar{m}_f^2) \big] 
	+ C_{{m}_f} e^{-2x}\:, \nonumber\\
  \bar{\sigma}^f_V(x) & = & 
  \frac{ 2(x+\bar{m}_f^2) - 1 + e^{-2(x+\bar{m}_f^2)} }
       { 2(x+\bar{m}_f^2)^2 }\:, \label{QuarkProp}
\end{eqnarray}
where ${\cal F}[x] = ( 1 - e^{-x})/{x}$,
$x = k^2/\lambda^2$, $\bar{\sigma}^f_{S} = \lambda \sigma^f_S$,
$\bar{\sigma}^f_V = \lambda^2 \sigma^f_V$, 
$\bar{m}_f = m_f / \lambda$,
$\lambda =$ 0.566~GeV, and $\epsilon = 10^{-4}$.
This dressed-quark propagator has {\em no} Lehmann representation and hence
describes the propagation of a confined quark. 
Furthermore, it reduces to a bare-fermion propagator with current-quark
mass $m_f$ for large, spacelike momenta, in accordance with perturbative
QCD. The parameters for the $u$-, $d$ and $s$-quark propagators were
determined in \Ref{Burden} by performing a $\chi^2$ fit to a range of
$\pi$- and $K$-meson observables. 

The final element is the $\rho$-meson BS amplitude $V_{\nu}(k;p)$ which
was modeled in \Ref{Pichowsky} as the sum of exponential and monopole
functions in the relative quark-antiquark momentum $k^2$.  The parameters
were determined by requiring the BS amplitude leads to the experimental
values for the $\rho \rightarrow \pi \pi$ and $\rho \rightarrow e^+ e^-$
decay widths.

\begin{figure}[h]
\centering{\ 
\mbox{\epsfig{figure=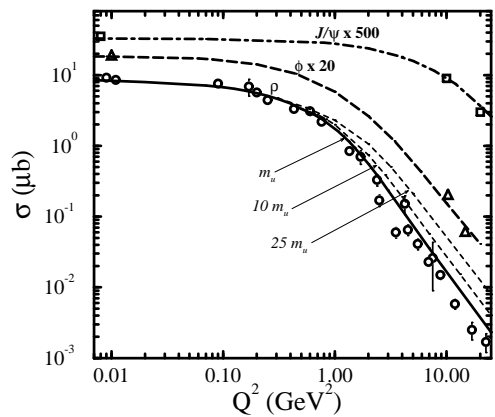,width=7.0cm}}}
\caption{
The $\rho^0$-, $\phi$-, and $J/\psi$-meson electroproduction cross
sections at $W =$ 15, 100 and 100 GeV, respectively. 
(The latter two results and corresponding data are rescaled by amounts
indicated.)   
The narrow dashed curves show $\rho$-meson electroproduction results
obtained using fictitious current-quark masses of $10$ and $25$
times larger than $m_u$ = 5.5~MeV. 
}
\label{Fig:TotalRho}
\end{figure}

Having determined the Schwinger functions in \Eq{tman} from
Dyson-Schwinger studies of low-energy hadron observables, and the
quark-nucleon pomeron-exchange interaction ${\cal G}_{\alpha}(W^2,t)$ from
meson-nucleon elastic scattering, one can use \Eqs{Current} and
(\ref{tman}) to calculate the $\rho$-meson electroproduction cross
section.
The result (solid curve in \Fig{Fig:TotalRho}) is in excellent agreement
with the data for all $q^2$.  
One important feature of the quark-nucleon interaction 
${\cal G}_{\alpha}(W^2,t)$ in \Eq{Current} is that it is independent of
$q^2$.   
Hence, all of the $q^2$ dependence in diffractive electroproduction
cross sections arises from the Schwinger functions in the quark-loop
integration of \Eq{tman}.
The $q^2$ dependence of the electroproduction cross section, shown in
\Fig{Fig:GRhoGluonX}, results from having employed elementary Schwinger
functions that describe low-energy meson observables such as the $\pi$-
and $K$-meson electromagnetic form factors.

In the study of \Ref{Pichowsky}, the role of the current-quark mass $m_u$
in determining the $q^2$ dependence of 
electroproduction was explored using fictitious current quark masses $m_f$
that are 10 and 25 times greater than $m_u = $ 5.5~MeV to recalculate the
$\rho$-meson electroproduction cross section.
The results are shown in \Fig{Fig:TotalRho}.
A comparison of these two curves to the result obtained using the correct
value of $m_u =$ 5.5~MeV, suggests two important features of diffractive
electroproduction. 
First, the magnitude of the photoproduction cross section ($q^2 = 0$) is
unaffected by changes to  $m_u$. 
This is because the quark-photon vertex $\Gamma_{\mu}^{\rm BC}(k,k')$
satisfies the Ward Identity which, when coupled with the normalized
$\rho$-meson BS amplitude, tightly constrains the magnitude of the cross
section at $q^2 = 0$.
Second, although all three curves converge in the photoproduction limit of
$q^2 \rightarrow 0$, they diverge significantly for larger values of
$q^2$.
Ultimately, these curves exhibit the same $q^{-4}$ behavior, 
but the onset of this asymptotic behavior is {\em determined} by the
magnitude of $m_u$; 
that is, a larger current-quark mass for the quark and antiquark inside
the produced vector meson postpones the onset of the asymptotic power-law
behavior until a larger value of $q^2$.
The results for $\phi$ and $J/\psi$ electroproduction obtained from the
model of \Ref{Pichowsky} are shown in \Fig{Fig:TotalRho}.  
Both the model calculation and experimental data exhibit the anticipated
behavior.  

The dependence of diffractive electroproduction on the current-quark mass
$m_f$ provides an explanation of the dramatic result that although the
$\rho$-meson electroproducton 
cross section is {\em two orders of magnitude} larger than that of the
$J/\psi$ meson near $q^2 = 0$, they are nearly equal for larger $q^2$.
This behavior is a result of having used the dressed-quark propagator
$S_f(k)$ from Dyson-Schwinger studies which evolve dynamically with the
quark momentum $k^2$.
For example, the behavior of the $u$-quark propagator $S_u(k)$ at small
momentum $k^2$ is characterized by constituent-quark-like mass scales
($\approx$ 330 MeV) while at larger momentum, it is characterized by the
current-quark mass $m_u$ = 5.5~MeV. 
The dynamical evolution of mass scales in the dressed-quark propagator
$S_f(k)$ is essential for the description of electroproduction at all
$q^2$. 

\section{Photo-meson transition amplitudes}
\begin{figure}[t]
\begin{center}
\epsfig{figure=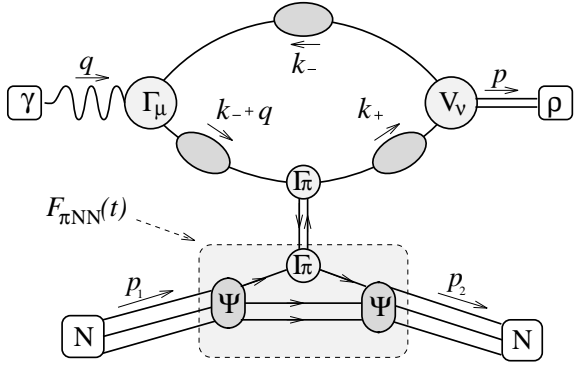,width=7.50cm}
\end{center}
\caption
{
The $\rho$-meson electroproduction current due to 
$t$-channel effective-$\pi$ exchange.
The shaded box represents the effective-$\pi NN$ form factor,
$F_{\pi NN}(t)$. 
}
\label{Fig:PiX}
\end{figure}
\begin{figure}[t]
\centering{ 
\mbox{\epsfig{figure=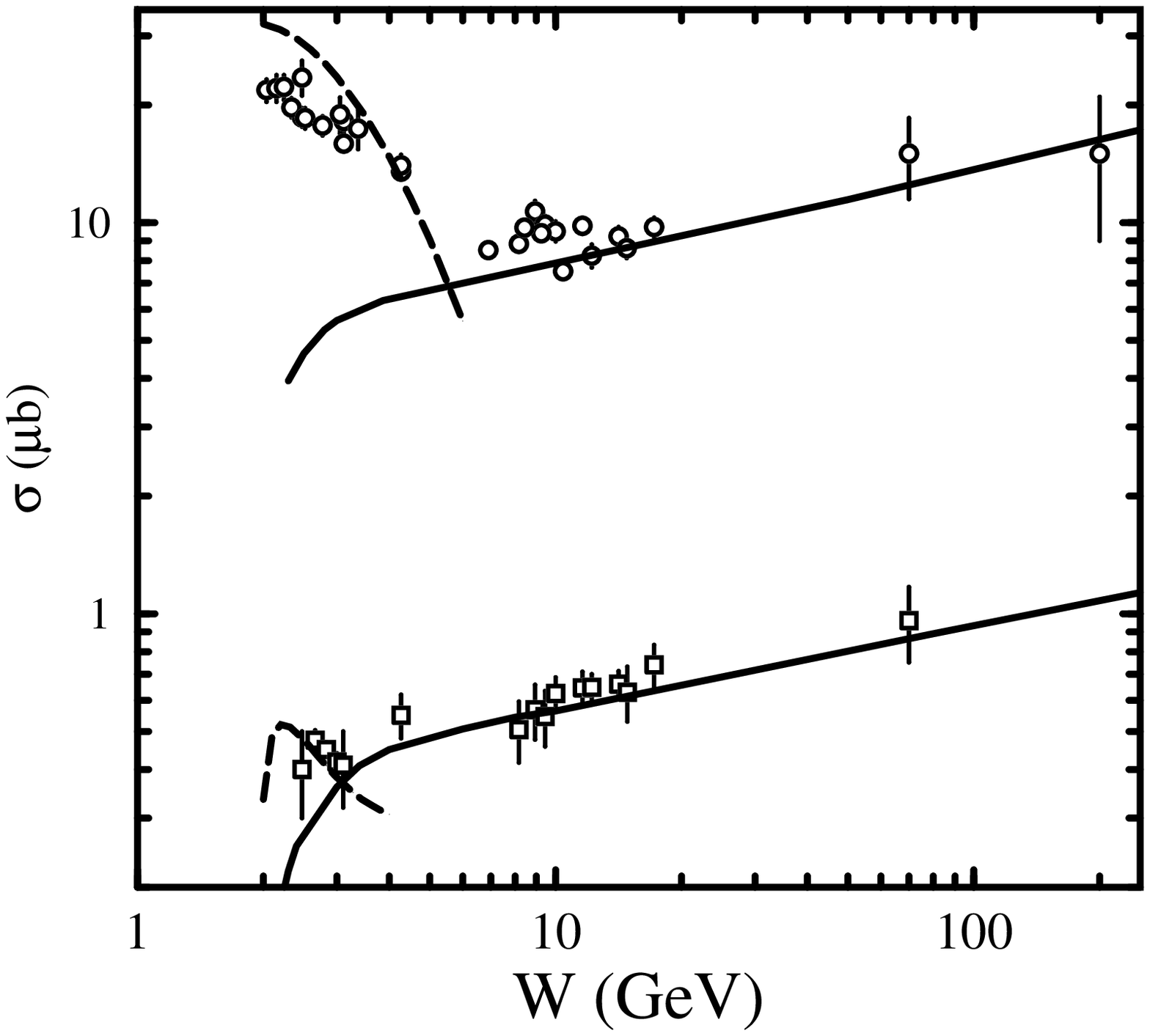,width=7.0cm}} \\
\vspace*{-0.55cm}
\mbox{\epsfig{figure=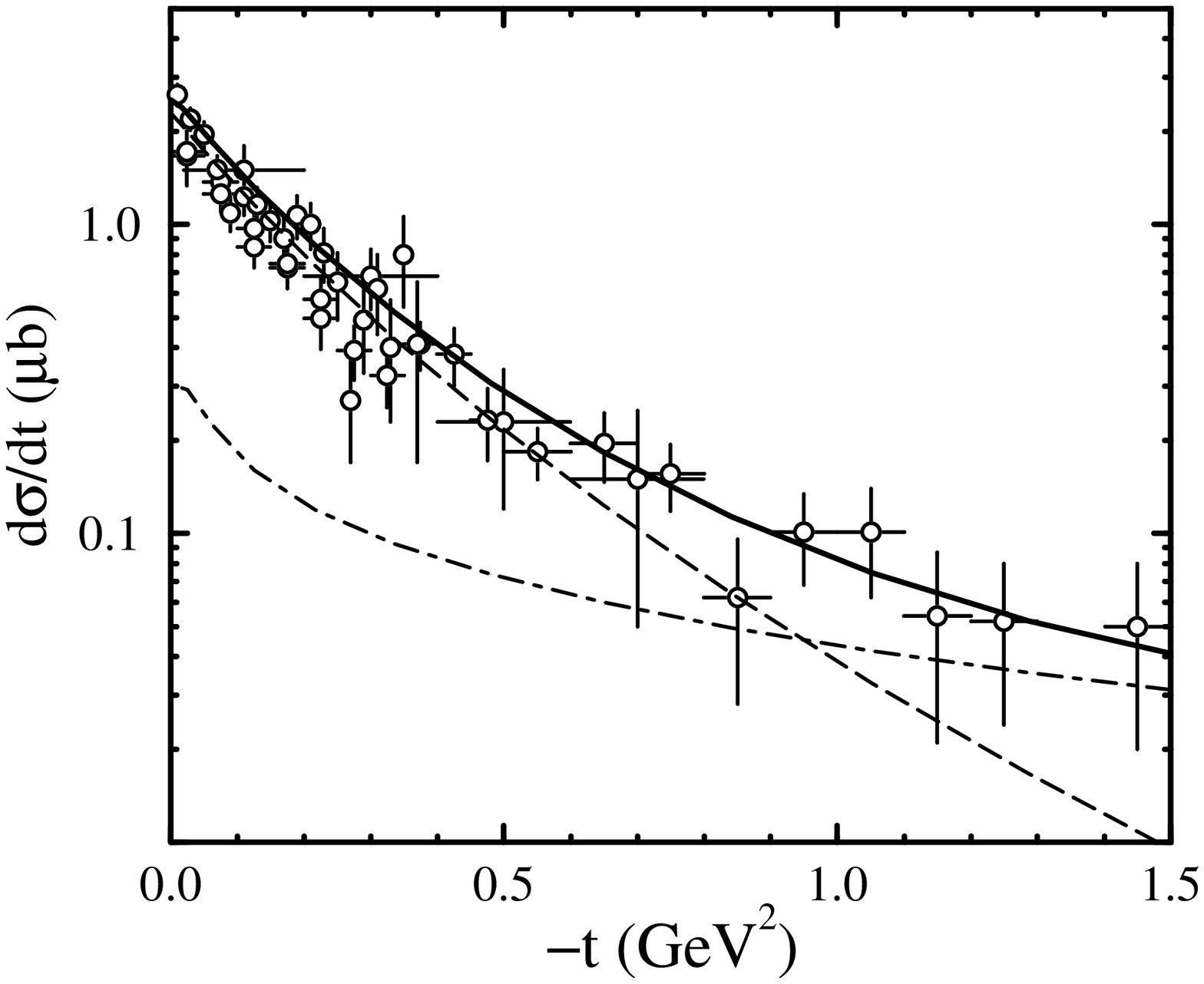,width=7.0cm}}}
\caption{
Top: Contributions to $\rho$- and $\phi$-meson photoproduction from pomeron
	exchange (solid) and meson exchanges (dashed).
Bottom: 
The differential cross section for $\phi$-meson photoproduction 
($q^2 =0$) 
for  $3.0 \leq W \leq$ 3.5~GeV. 
The contributions due to the pomeron exchange (dashed), and 
$\pi$ and $\eta$ exchanges (dot-dashed) and their sum (solid) are
shown.
}
\label{Fig:W}
\label{Fig:dsigt3}
\end{figure}

%
At lower energies, both pomeron and correlated-quark exchanges are
important.  The energy at which pomeron exchange and effective-meson
exchanges each contribute about a half of the photoproduction cross
section is $W\approx$ 6~GeV for $\rho$ mesons and $W\approx$ 3~GeV for
$\phi$ mesons. (See \Fig{Fig:W}.)
These are incorporated into the pomeron-exchange model by a reorganization
of the nonlocal interactions between quarks into sums of exchanges of
effective fields with mesonic quantum numbers, referred to as 
effective-meson fields.  An account of how this procedure is 
carried out in theory and practice is given in \Ref{Tandy}.

As an example, consider the contribution to $\rho$-meson electroproduction
due to the $t$-channel exchange of an effective ``$\pi$ meson''.  
Applying the techniques described in \Ref{Tandy}, one obtains the
electroproduction current matrix element:
\begin{eqnarray}
\lefteqn{
\langle p \lambda_{\rho} ; p_2 m_s^{\prime} | J_{\mu}(q) 
| p_1 m_s \rangle =  \Lambda_{\mu \nu}(q,p) \,
\varepsilon_{\nu}(p;\lambda_{\rho})  }
\nn\ \\ 
& & \times  
\; 
\frac{1}{m_{\pi}^2 - t} \; 
\, \bar{u}(p_2,m_s^{\prime}) \;
 \gamma_{5} F_{\pi N N}(t) \; 
u(p_1,m_s) 
,
\label{JPiX}
\end{eqnarray}
where 
$\Lambda_{\mu \nu}(q,p)$ is the photon-$\rho$-meson transition amplitude
due to the exchange of an effective $\pi$ meson (denoted in \Fig{Fig:PiX} 
by a quark loop) and 
$F_{\pi N N}(t) = ( 1 - t/\Lambda^2)^{-1}$ with $\Lambda \approx$ 1~GeV
is the effective-$\pi$-$NN$ form factor (denoted by a shaded box in
\Fig{Fig:PiX}). 

\begin{figure}[ht]
\begin{center}
\epsfig{figure=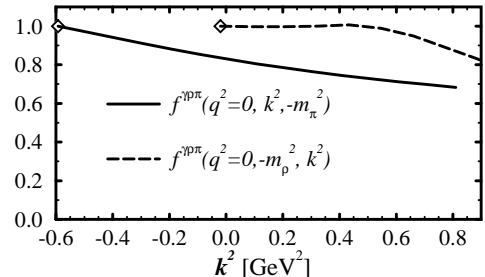,width=7.50cm}\\
\vspace*{-2.7cm}
\end{center}
\caption
{
The off-shell $\gamma\rho\pi$ transition form factor is shown for the 
cases when either the $\rho$ or $\pi$ meson is off shell.
Diamonds indicate the point at which all three particles are on shell.
In a calculation of $\rho$-meson photoproduction the $\rho$-meson is
on-shell so that only the latter form factor (dashed curve) is needed.
}
\label{Fig:GRhoPiFF}
\end{figure}

The photo-$\rho$-meson transition amplitude $\Lambda_{\mu \nu}(q,p)$ is
given by
\begin{eqnarray} 
\lefteqn{
\Lambda_{\mu \nu}(q,p)
=  \frac{e_0}{\sqrt{Z_{\pi}(t)}} 
\;  {\rm tr} \int\!\! \frac{{\rm d}^4 k}{(2\pi)^2} 
	\; S_u(k_{-})  \Gamma_{\mu}(k_{-},k_{-}\+q) 
} \nn\ \\
& & \times \; 
S_u(k_{-}\+q) V_{\nu}(k;p)
S_u(k_{+}) {\Gamma_{\pi}(k+\sfrac{1}{2}q;-t)}
,\label{GRhoPi}
\end{eqnarray}
where $\Gamma_{\pi}(k;p)$ is the on-shell pion BS amplitude, 
and $Z_{\pi}(t)$ is the factor appearing in the effective-meson
``propagator'' 
$\Delta_{\rm eff}(t) = Z^{-1}_{\pi}(t) \; (m_{\pi}^2 - t)^{-1}$ 
which arises from the nonlocal nature of the effective pion.
The factor $Z_{\pi}(t)$ is calculated in a straight-forward
manner from the $\pi$-meson BS amplitude $\Gamma_{\pi}(k;p)$ and the
dressed-quark propagator $S_f(k)$, as described in \Ref{Tandy}.
Considerations of parity and Lorentz covariance allow one to rewrite the 
amplitude as
\begin{equation}
\Lambda_{\mu \nu}(q,p) = 
\frac{e_0}{m_{\rho}} \; g_{\gamma\rho\pi} \;
  \epsilon_{\mu \nu \alpha \beta} q_{\alpha} p_{\beta}
\; f^{\gamma\rho\pi}(q^2,p^2,-t)
,
\label{g_grp:Def}
\end{equation}
where $g_{\gamma\rho\pi}$ is the $\rho\rightarrow \gamma\pi$ decay
constant, $\epsilon_{\mu \nu \alpha \beta}$ is the usual Levi-Cevita
tensor, 
and $f^{\gamma\rho\pi}(q^2,p^2,-t)$ is the $\gamma\rho\pi$-transition form
factor, defined so that
when all three particles are on their mass shell, 
$f^{\gamma\rho\pi}(q^2=0,p^2=-m_{\rho}^2,-t=-m_{\pi}^2) = 1$. 
All of the elements required to calculate \Eq{JPiX} are known.
The resulting photo-$\rho$-meson transition form factor is shown in 
\Fig{Fig:GRhoPiFF} for a range of off-shell $\pi$- and $\rho$-meson
momenta.

The slow fall off of the form factors shown in \Fig{Fig:GRhoPiFF} is
typical of photo-meson transition amplitudes.  
It is a result of the fact that the factor $\sqrt{Z(t)}$ decreases with
$-t$, which tends to weaken and partially counteract the rapid fall off
usually observed in spacelike transition form factors when all mesons are
on shell.  
This behavior has also been observed in studies of other photo-meson
transition amplitudes, 
such as $\gamma \gamma \pi$ and 
$\gamma \pi \rightarrow \pi \pi$ \cite{Frank}.

With these meson-transition form factors, one can use the quark-nucleon
pomeron-exchange model to calculate $\rho$- and $\phi$-meson
electroproduction at all energies and $q^2$.  This work is currently in
progress.  However, shown in the top of \Fig{Fig:W} are the cross sections
for $\rho$- and $\phi$-meson production with 
$f^{\gamma\rho\pi}(q^2=0,p^2=-m_{\rho}^2,-t) = 1$.
Hence, these predictions should overestimate the contributions from meson
exchanges.  
In the bottom plot of \Fig{Fig:W}, are results for the $\phi$-meson
differential cross section at $W =$ 3~GeV.
At moderate $t$, the meson-exchange contributions (from $\pi$ and $\eta$
exchange) overwhelm that of pomeron-exchange.  At lower energies,
meson exchanges become increasingly more important and one must include
them (and their respective transition form factors) to obtain good
agreement with the data, even at small $t$.

\section{Conclusion}
I have given a brief outline of how quark dynamics and correlated-quark
exchanges may be explored in diffractive electroproduction using a quantum
field theoretic framework based on the Dyson-Schwinger equations of QCD. 

In numerical studies of the Dyson-Schwinger equations, both perturbative
and nonperturbative aspects of QCD are manifest in the solutions obtained
for the elementary Schwinger functions, such as the dressed-quark
propagator and meson Bethe-Salpeter amplitudes.
In phenomenological applications, such as the one
described here, one explores the consequences of employing such dressed
Schwinger functions and their effect on experimental observables.  
In this way, one is able to probe the underlying dyamics of quarks and
gluons involved in exclusive processes and further improve our
understanding of QCD.

\section*{Acknowledgments}

\vspace*{-0.3cm}
This work is supported by the U.S. Department of Energy
under Contracts DE-AC05-84ER40150 and DE-FG05-92ER40750 and 
the Florida State University Supercomputer Computations Research Institute
which is partially funded by the Department of Energy through 
Contract DE-FC05-85ER25000.
\vspace*{-0.4cm}

\end{document}